\documentclass{article}

\usepackage{PRIMEarxiv}

\usepackage{url}            % simple URL typesetting
\usepackage{booktabs}       % professional-quality tables
\usepackage{nicefrac}       % compact symbols for 1/2, etc.
\usepackage{microtype}      % microtypography
\usepackage{fancyhdr}       % header
\usepackage{graphicx}       % graphics
\usepackage{qcircuit}
\usepackage{subcaption}
\usepackage{physics}
\usepackage{authblk}

\title{An Annealing-based Quantum Neural Network with Discrete Wavelet Transforms for Enhanced Air Pollution Nowcasting
}

\author[1,2]{Otto Menegasso Pires\thanks{Corresponding Author: otto.pires@fieb.org.br}}
\author[1,2]{André Saimon}
\author[6]{Daniela Herrmann}
\author[3,4,5]{Prashant Kumar}
\author[1,2]{Marcelo Moret}
\author[1,2,3,4,5]{Erick Giovani Sperandio Nascimento}

\affil[1]{QuIIN - Quantum Industrial Innovation, SENAI CIMATEC University, Avenida Orlando Gomes, 1845, Salvador, 41650-010, Bahia, Brazil}
\affil[2]{Latin-American Quantum Computing Center, SENAI CIMATEC University, Avenida Orlando Gomes, 1845, Salvador, 41650-010, Bahia, Brazil}
\affil[3]{Surrey Institute for People-Centered Artificial Intelligence, Faculty of Engineering and Physical Sciences, University of Surrey, Guildford, GU2 7HX, UK}
\affil[4]{Global Centre for Clean Research (GCARE), School of Sustainability, Civil and Environmental Engineering and Physical Sciences, University of Surrey, Guildford, GU2 7HX, UK}
\affil[5]{Institute for Sustainability, University of Surrey, Guildford, GU2 7HX, UK}
\affil[6]{Dynex, QEnterprises AG, Blegistrasse 3, Baar, 6340, Switzerland}

\begin{document}
\maketitle

\begin{abstract}
Air pollution poses a significant threat to public health, exacerbated by the rapid industrialization and urbanization of modern society. While traditional deep learning offers predictive capabilities, recent advancements in quantum machine learning (QML) present a compelling alternative. Here, we introduce a scalable hybrid framework that integrates variational quantum circuits with quantum annealing. Leveraging a neuromorphic quantum computing platform to transcend the qubit limitations of the NISQ era, we model particulate matter (PM$_{2.5}$) dynamics using high-density urban data from the UK. We demonstrate that our Quantum Neural Networks (QNNs) exhibit an intrinsic resilience to input volatility, significantly outperforming classical baselines (MLP and LSTM) on raw, high-entropy environmental data. Furthermore, we successfully scaled the model to 172 qubits, observing a synergistic enhancement when coupled with wavelet decomposition that achieves state-of-the-art predictive fidelity. These findings establish a critical benchmark for quantum utility, proving that scalable, neuromorphic quantum machine learning can now address complex, non-linear real-world problems.
\end{abstract}

% keywords can be removed

\section{Introduction}
%%%%%%%%%%%
% SETTING %
%%%%%%%%%%%
Urbanization is accelerating, with 55\% of the global population living in cities by 2018~(\cite{urbanization}) and projections reaching 68\% by 2050~(\cite{UN-urban}). 
Anthropogenic activities, particularly those related to industrial processes and urbanization, stand as primary culprits in the emission of hazardous substances into our atmosphere~(\cite{climate-watch}).
As a result, air pollution has become a major global health crisis, contributing to millions of deaths each year and a significant loss of healthy life years, rivaling the health risks posed by poor diets and smoking~(\cite{WHO-air-guidelines}).
Urban areas are particularly affected, with industrial activity and road traffic releasing pollutants such as particulate matter (PM), nitrogen oxides (NOx), carbon monoxide (CO), and volatile organic compounds (VOCs)~(\cite{2022Kitagawa}). Among these, fine particulate matter (PM$_{2.5}$) poses the greatest threat, as exposure has been linked to severe respiratory and cardiovascular diseases, including asthma, chronic bronchitis, stroke, and lung cancer~(\cite{WHO-air-health-impacts}). Notably, research indicates that no level of PM exposure is safe, as even low concentrations can harm human health~(\cite{WHO-effects-pm}).
To address this growing crisis, the World Health Organization (WHO) calls for stricter air quality management, improved monitoring of PM levels, and measures to reduce pollution. Without urgent action, the health impacts of air pollution will continue to escalate, particularly in densely populated urban regions.

%%%%%%%%%%%%%%%%%%%%%%%%%%%%%%%%%%%%%%%%%%%%%%%%%
% COMPUTATIONAL INTELLIGENCE FOR PM FORECASTING %
%%%%%%%%%%%%%%%%%%%%%%%%%%%%%%%%%%%%%%%%%%%%%%%%%
Particulate air pollution can be reduced using current technologies.
Deep learning (DL) has emerged as the dominant paradigm for analyzing the non-linear spatiotemporal dynamics of air pollution, surpassing traditional chemistry-transport and shallow statistical models~(\cite{2020Liao}). To handle the non-stationary nature of environmental data, state-of-the-art approaches increasingly rely on hybridization strategies. These methods typically couple data decomposition techniques, such as Empirical Mode Decomposition~(\cite{2021Huang}) or Discrete Wavelet Transforms~(\cite{2022Galvao}), with complex neural architectures, combining Convolutional Neural Networks (CNNs) for spatial feature extraction with Recurrent Neural Networks (RNNs) for temporal modeling~(\cite{2023Mendez}). While diverse implementations, such as auto-encoder ensembles~(\cite{2020Zhang}), physics-informed enhancements of the CMAQ model~(\cite{2021ChangHoi}) and the integration of Random Forest to handle incomplete data~(\cite{2023Guo}), have consistently improved forecasting accuracy~(\cite{2022Sharma}), they achieve this performance at the cost of exponentially increased computational complexity and reduced energy efficiency~(\cite{2021Zaini}).

%%%%%%%%%%%%%%%%%%%%%
% THE ISSUE WITH AI %
%%%%%%%%%%%%%%%%%%%%%

Although artificial intelligence holds promise for advancing sustainability goals, the high energy requirements for training and tuning models remain a pressing concern~(\cite{2021Wynsberghe}). The computational demands of sophisticated hybrid deep learning models substantially increase the time required for model training~(\cite{2021Zaini}), resulting in notable energy consumption and environmental impact~(\cite{2023Islam}).
Deep learning research, in particular, contributes to a growing carbon footprint due to these intensive computational needs~(\cite{2020Schwartz}). Furthermore, model refinement processes often require even more energy than initial training, amplifying the environmental cost of AI development~(\cite{2021Wynsberghe}). To address these challenges, researchers are increasingly focused on data-efficient methods that can achieve competitive performance with reduced training data or fewer optimization steps, helping to lessen AI's ecological impact.

%%%%%%%%%%%%%%%%%%%%%%%%%
% QML AS AN ALTERNATIVE %
%%%%%%%%%%%%%%%%%%%%%%%%%

Quantum Machine Learning (QML) presents a compelling alternative to address these sustainability concerns. By exploiting quantum mechanical properties such as superposition and entanglement, QML algorithms can theoretically navigate high-dimensional feature spaces more efficiently than classical counterparts~(\cite{2022Avramouli}), potentially reducing the data and training epochs required for accurate modeling~(\cite{2023Enriquez}).
However, a clear quantum advantage in environmental forecasting remains elusive. While early hybrid architectures, combining classical optimization with Quantum Long Short-Term Memory (QLSTM) networks, have demonstrated predictive capability in air quality tasks~(\cite{2024Li}), they are frequently bottlenecked by the noise of current hardware~(\cite{2022Cerezo}) and the challenge of encoding classical data into quantum states~(\cite{2022Houssein}). 

%%%%%%%%%%%%%%%%%%%%%%%%%%%%%%
% NOVELTIES AND CONTRIBUTION %
%%%%%%%%%%%%%%%%%%%%%%%%%%%%%%
Here, we bridge this gap by introducing a scalable Quantum Neural Network (QNN) framework for PM$_{2.5}$ nowcasting. By integrating multi-level wavelet transforms to capture non-linear temporal dynamics, our approach demonstrates superior predictive performance compared to classical MLP and LSTM baselines, yielding higher Pearson correlations and reduced error rates.
Crucially, we address the scalability limits of current hardware by validating our architecture on  the DynexQML~(\cite{Dynex}) framework, that enables quantum neural networks through established quantum circuit translation approaches, based on theoretical work in quantum annealing formulations~(\cite{2014Warren}). 

The main contributions of this paper are listed below:
\begin{itemize}
    \item We introduce a scalable hybrid Quantum Neural Network (QNN) framework that couples variational quantum circuits with neuromorphic quantum annealing to transcend physical NISQ-era hardware limitations.
    \item We demonstrate that the QNN architecture exhibits an intrinsic resilience to input volatility, substantially outperforming classical MLP and LSTM baselines on raw, unprocessed temporal environmental data.
    \item We systematically validate the architecture's scalability by successfully simulating circuits of up to 172 qubits, observing state-of-the-art predictive fidelity when coupled with multi-level Discrete Wavelet Transforms.
    \item We observe a non-linear relationship between model complexity and performance, where increasing parameters such as qubit count or lookback period did not yield proportional improvements.
\end{itemize}

The organization of this paper is as follows: Section \ref{sec:methodology} explicitly defines the proposed quantum neural network methodology and the neuromorphic translation framework. Section \ref{sec:experimental} details the experimental setup, data processing pipelines, and baseline classical models. Section \ref{sec:results} presents the comparative evaluation and benchmarking results. Finally, Section \ref{sec:conclusion} concludes the paper and discusses future work.

\section{Proposed Methodology}
\label{sec:methodology}

\subsection{Quantum Neural Network}
Quantum neural networks (QNNs) represent a novel approach to machine learning, blending classical data processing with the power of quantum computing. Despite their classical underpinnings, QNNs are considered ``purely quantum'' models because their execution relies solely on classical computing for circuit preparation and statistical analysis~(\cite{2023Combarro}). These networks fall under the umbrella of Variational Quantum Algorithms, employing Parameterized Quantum Circuits (PQCs), also known as ansätze, which are trained using classical optimization techniques.

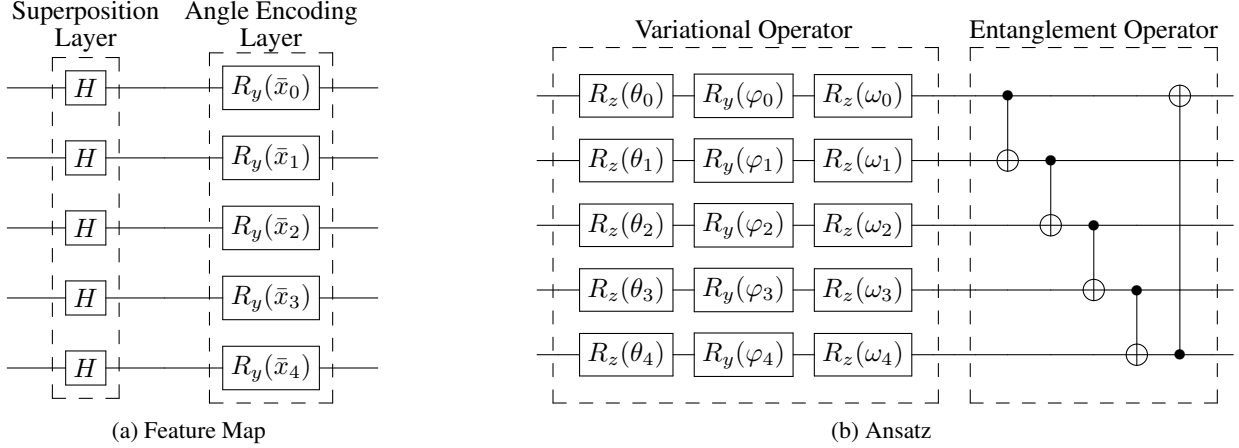
\begin{figure}
\centering
    \begin{subfigure}{0.3\textwidth}
        \centerline{
            \Qcircuit @C=2.2em @R=1em {
            & \mbox{Superposition}&& \mbox{Angle Encoding}& \\
            & \mbox{Layer}        && \mbox{Layer}         & \\
            & \gate{H}            &\qw& \gate{R_y(\bar{x}_0)} &\qw\\
            & \gate{H}            &\qw& \gate{R_y(\bar{x}_1)} &\qw\\
            & \gate{H}            &\qw& \gate{R_y(\bar{x}_2)} &\qw\\
            & \gate{H}            &\qw& \gate{R_y(\bar{x}_3)} &\qw\\
            & \gate{H}            &\qw& \gate{R_y(\bar{x}_4)} &\qw \gategroup{3}{2}{7}{2}{1em}{--} \gategroup{3}{4}{7}{4}{1em}{--}
            }
        }
        \caption{Feature Map}
        \label{fig:data-encoding-layer}
    \end{subfigure}
    \hfill
    \begin{subfigure}{0.59\textwidth}
        \centerline{
        \Qcircuit @C=0.8em @R=0.8em {
        &&&\mbox{Variational Operator}&&&&&&&\mbox{Entanglement Operator}&&& \\
        &&&&&&&&&&&&&& \\
        & \qw & \gate{R_z(\theta_0)} & \gate{R_y(\varphi_0)} & \gate{R_z(\omega_0)} &\qw&\qw&\qw & \ctrl{1} & \qw      & \qw      & \qw      & \targ     & \qw & \qw   \\
        & \qw & \gate{R_z(\theta_1)} & \gate{R_y(\varphi_1)} & \gate{R_z(\omega_1)} &\qw&\qw&\qw & \targ    & \ctrl{1} & \qw      & \qw      & \qw       & \qw & \qw  \\
        & \qw & \gate{R_z(\theta_2)} & \gate{R_y(\varphi_2)} & \gate{R_z(\omega_2)} &\qw&\qw&\qw & \qw      & \targ    & \ctrl{1} & \qw      & \qw       & \qw & \qw  \\
        & \qw & \gate{R_z(\theta_3)} & \gate{R_y(\varphi_3)} & \gate{R_z(\omega_3)} &\qw&\qw&\qw & \qw      & \qw      & \targ    & \ctrl{1} & \qw       & \qw & \qw \\
        & \qw & \gate{R_z(\theta_4)} & \gate{R_y(\varphi_4)} & \gate{R_z(\omega_4)} &\qw&\qw&\qw & \qw      & \qw      & \qw      & \targ    & \ctrl{-4} & \qw & \qw \gategroup{2}{2}{8}{6}{0.4em}{--} \gategroup{2}{8}{8}{14}{0.4em}{--}\\
        &&&&&&&&&&&&&&
        }
    }
        \caption{Ansatz}
        \label{fig:strong-layer}
    \end{subfigure}
\caption{The feature map (a) is composed of a Superposition Layer and an Angle Encoding Layer, allocating one qubit for each feature within the dataset. During the encoding procedure, the circuit applies a rotation gate to each qubit $j$, with the rotation angle dictated by the parameterized value $\bar{x}_j$ corresponding to the normalized specific feature. The ansatz (b) of our quantum neural network model is defined by a variational operator parametrized by the vectors $\boldsymbol{\theta}, \boldsymbol{\varphi}, \boldsymbol{\omega}$ followed by a circular entanglement scheme as our entanglement operator. We define the circuit depth as the number of repetitions of the ansatz. Consequently, a model with depth-2 concatenates two identical blocks, each containing one variational stage and one entanglement stage, resulting in a total of four operator layers.}
\label{fig:QNN}

\end{figure}

The behavior of QNNs mirrors that of classical neural networks, comprising three key stages: Data Preparation, Model Training, and Model Output. The Data Preparation stage involves encoding classical data into a quantum state. This is done in two steps, as shown in Figure \ref{fig:data-encoding-layer}: first, a Superposition Layer applies Hadamard gates to all $N$ qubits, one for each feature, to create a superposition state. Second, an Angle Encoding Layer applies a single-qubit Y-rotation gate, $R_y(\bar{x}_j)$, to each qubit, where the rotation angle is directly determined by the normalized input feature $\bar{x}_j$.

Following data preparation, the model training occurs within the framework of its ansatz, depicted in Figure \ref{fig:strong-layer}. The ansatz consists of $L$ repeated layers. Each layer contains a Variational Operator, which applies a sequence of single-qubit $R_z(\theta)R_y(\varphi)R_z(\omega)$ gates with trainable angles, and an Entanglement Operator, which uses a circular entanglement scheme with CNOT gates applied between adjacent qubits. The number of layers $L$ can be adjusted to influence the model's learning capacity.

Finally, the Model Output stage generates predictions. This involves performing a computational basis measurement on all $N$ qubits to obtain the expectation value for each qubit. These expectation values are then fed into a classical post-processing layer consisting of fully connected neurons with a ReLU activation function. This classical layer performs a linear transformation of the quantum output to produce the final predictions for a specific forecast horizon, which is equal to the number of nodes in the layer.

\subsection{DynexQML Framework and Circuit Translation}
Our study employs DynexQML, a novel framework built on the Dynex neuromorphic quantum computing platform ~(\cite{Dynex}). This platform transforms conventional Graphics Processing Units (GPUs) into quantum annealing processors through memristive emulation, which simulates quantum superposition states. This approach facilitates quantum annealing computations at scale without the constraints of traditional quantum hardware.

\begin{figure}
    \centering
    \includegraphics[width=\linewidth]{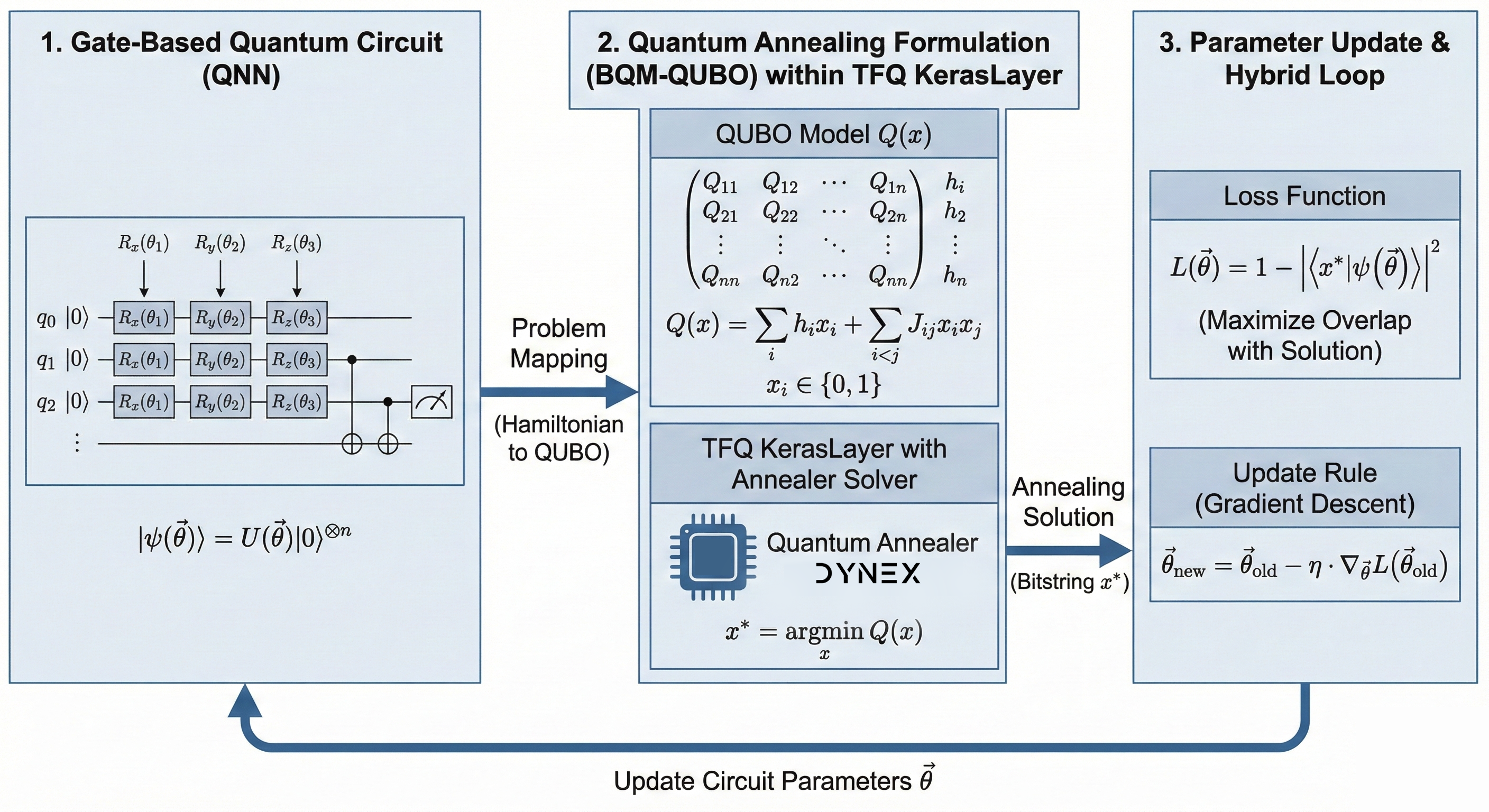}
    \caption{Schematic of the DynexQML Framework for Hybrid Quantum-Classical Machine Learning. The framework integrates a gate-based QNN with neuromorphic quantum annealing emulation for scalable training. Steps: 1. The process begins with a gate-based quantum circuit; 2. This circuit is mapped to a QUBO problem, represented by a Hamiltonian $Q(x)$. The problem is embedded within a QKerasLayer and solved on the Dynex neuromorphic platform, which emulates quantum annealing to find the ground state $x^\ast=\arg\min_xQ(x)$; 3. The annealing solution (bitstring $x^\ast$) is used to compute a loss function $L(\theta)$ based on the overlap with the target state. Circuit parameters $\theta$ are then updated via gradient descent in a hybrid loop, iteratively refining the quantum model. This architecture enables the training of large-scale quantum models by leveraging the scalability of neuromorphic annealing for optimization tasks that are typically intractable on near-term gate-based hardware.}
    \label{fig:qkeraslayer}
\end{figure}

The framework's core is the QKerasLayer, which introduces a distinct learning paradigm by encoding and training models using quantum annealing computations (see Fig. \ref{fig:qkeraslayer}, encoding stage). This layer achieves learning and optimization through epochal quantum sampling, where discrete quantum state sampling is performed at the boundaries of each training epoch, rather than relying on continuous quantum evolution. This protocol is designed to reduce potential quantum decoherence by limiting the system's exposure time and enhances scalability through distributed quantum annealing operations.

To execute a computation, the complete quantum circuit is encoded into a time-independent Hamiltonian, $H_{circuit}$, which represents the problem to be solved:
\begin{equation}
   H_{circuit} = \sum_i \alpha_i \hat{H}_i^{(gate)} + \sum_j \beta_j \hat{H}_j^{(constraint)} 
\end{equation}
In this formulation, the $\hat{H}_i^{(gate)}$ terms represent the quantum operations, while the $\hat{H}_j^{(constraint)}$ terms enforce the circuit's topology. As shown in Fig. \ref{fig:qkeraslayer}, the layer encodes both input data ($x_{input}$) and trainable parameters ($\theta_{weights}$) directly into this quantum formulation, where input data is mapped to quantum amplitudes and the neural weights are encoded as quantum parameters within the Hamiltonian structure.

The learning process is framed as an energy minimization problem, where the quantum annealing computation seeks the ground state configuration, $\ket{\Psi_{ground}}$, of the problem Hamiltonian:
\begin{equation}
  \ket{\Psi_{ground}} = \arg\min_\Psi \bra{\Psi}H_{circuit}\ket{\Psi}  
\end{equation}
This ground state represents the optimal quantum configuration for a given input-weight combination. This minimum is found by following the quantum adiabatic evolution protocol.

A key feature of the QKerasLayer is its customized gradient computation. Rather than computing gradients in the traditional gate parameter space, updates are derived from the quantum energy space. The energy-based gradient for a given parameter $\theta_i$ is computed using a finite difference approximation:
\begin{equation}
  \frac{\partial E_{quantum}}{\partial \theta_i} \approx \frac{E(\theta_i + \epsilon) - E(\theta_i - \epsilon)}{2\epsilon}  
\end{equation}
where $E(\theta_i)$ is the quantum energy expectation value for that parameter configuration. The quantum parameters are then updated using this energy gradient, forming a feedback loop as illustrated in Fig. \ref{fig:qkeraslayer}:
\begin{equation}
  \theta_{quantum}^{(t+1)} = \theta_{quantum}^{(t)} - \eta \nabla_\theta E_{quantum}(\theta^{(t)})  
\end{equation}
where $\eta$ is the learning rate and $E_{quantum}$ represents the functional quantum energy.

Upon completion of the annealing process, quantum observables are extracted from the resulting ground state (Fig. \ref{fig:qkeraslayer} measurement stage),
\begin{equation}
  \langle O_i \rangle = \bra{\Psi_{ground}}\hat{O}_i\ket{\Psi_{ground}}
\end{equation}
These observables typically represent Pauli-Z measurements ($\hat{O}_i = \sigma_i^z$). The complete set of measurements on the system yields a feature vector composed of these expectation values, $[\langle Z_0 \rangle, \langle Z_1 \rangle, \ldots, \langle Z_{n-1} \rangle]$. This vector serves as the output of the quantum layer and is passed as input to subsequent classical neural network layers for further processing, completing the framework outlined in Fig. \ref{fig:qkeraslayer}.

\subsection{Quantum Bridge: Gate-to-Annealing Translation}
The DynexQML framework's capacity to execute deep quantum circuit algorithms relies upon a systematic translation of gate-based operations into an energy-based optimization framework. This ``Quantum Bridge'' protocol maps the unitary logic of a quantum circuit onto a Quadratic Unconstrained Binary Optimization (QUBO) problem. The theoretical foundation for this mapping, which demonstrates how arbitrary quantum circuits can be cast as time-independent Hamiltonians, follows established methodologies for quantum annealing ~(\cite{2014Warren}). Consequently, the emulated neuromorphic hardware finds the ground state of this Hamiltonian, which mathematically corresponds to the output state of the original gate-based computation.

\begin{table*}[h]
    \centering
    % Section for Basic Gates
    \begin{tabular}{l l c}
        \toprule
        \textbf{Basic Gate} & \textbf{Logical Qubits} & \textbf{Additional Qubits ($\delta_i$)} \\
        \midrule
        Pauli-X, Y, Z & 1 logical qubit & 1 \\
        Hadamard & 1 logical qubit & 3 \\
        T Gate & 1 logical qubit & 1 \\
        CNOT & 2 logical qubits & 2 \\
        Toffoli & 3 logical qubits & 3 \\
        Fredkin & 3 logical qubits & 4 \\
        CRX, CRY, CRZ & 2 logical qubits & 1 \\
        \midrule[\heavyrulewidth] 
        \textbf{Composite Operation} & \textbf{Input Parameters} & \textbf{Additional Qubit Overhead ($\delta$)} \\
        \midrule
        Quantum Fourier Transform (QFT) & $n$ logical qubits & $\frac{n^2}{2} + n$ \\
        Grover Operator & $n$ logical qubits & $2n + \log_2(n)$ \\
        Quantum Phase Estimation (QPE) & $m$ precision qubits, & $m \times (2^n - 1) + n$ \\
         & $n$ system qubits & \\
        \bottomrule
    \end{tabular}
    \caption{The number of additional physical annealing variables ($\delta_i$) required to represent gate-based operations as Hamiltonian constraints, divided into basic gates and scalable composite operations.}
    \label{tab:combined_scaling}
\end{table*}

A critical aspect of this translation is the resource overhead. It is imperative to distinguish between the \textit{logical qubits} defining the width of the quantum circuit (the feature dimension) and the \textit{annealing variables} (physical representational qubits) required to map the QUBO problem onto the neuromorphic device. Because each quantum gate imposes a mathematical constraint that must be embedded into the energy landscape, the translation incurs an additive overhead in the number of required annealing variables. The total number of annealing qubits ($Q_{total}$) required to represent a circuit with $n_{logical}$ logical qubits and $G$ distinct gate operations is expressed as:
\begin{equation}
   Q_{total} = n_{logical} + \sum_{i=1}^{G} \delta_i + n_{ancilla} 
\end{equation}
where $n_{logical}$ is the original circuit width, $\delta_i$ represents the localized variable overhead for gate $i$, and $n_{ancilla}$ accounts for any global ancilla variables required for connectivity or readout constraints. The gate-specific overhead $\delta_i$ scales depending on the complexity of the quantum operation, as detailed in Table \ref{tab:combined_scaling}.

To contextualize this scaling, we examine the largest architecture evaluated in this study: the 172-logical-qubit QNN utilized for processing the Wavelet-transformed environmental data (Lookback 2). While the circuit was designed as a 172-qubit system in reference to the quantum feature space, the actual QUBO optimization problem solved by the neuromorphic platform is significantly larger due to gate translation overhead. Assuming the Depth-4 architecture comprising parameterized single-qubit rotations and linear CNOT entanglement:
\begin{itemize}
    \item \textbf{Base State:} $n_{logical} = 172$ baseline variables.
    \item \textbf{Variational Layers:} 4 layers $\times$ 172 single-qubit rotations yields 688 gates. Assuming an average $\delta_i = 1$ per parameterized rotation, the overhead is 688 additional annealing variables.
    \item \textbf{Entanglement Layers:} 4 layers $\times$ 171 linear CNOT gates yields 684 two-qubit operations. Given $\delta_i = 2$ for CNOTs (Table \ref{tab:combined_scaling}), the overhead is 1,368 additional annealing variables.
\end{itemize}
Summing these components, the transformation yields:
\begin{equation}
    Q_{total} \approx 172 + 688 + 1368 = 2,228 \text{ annealing variables}
\end{equation}
Therefore, the 172-qubit QNN translates into an unconstrained binary optimization problem spanning over 2,200 interconnected variables. This translation exemplifies the critical necessity of the Dynex framework: representing an energy landscape of this dimensionality natively exceeds the coherence and connectivity limits of current physical gate-based QPUs, requiring the massive variable-capacity of neuromorphic emulation to find the ground state.

\section{Experimental Setup}
\label{sec:experimental}

\subsection{iSCAPE Dataset Characteristics and Feature Engineering}
The empirical validation of our architectures utilized high-density urban air quality data sourced from the iSCAPE (Improving the Smart Control of Air Pollution in Europe) project public platform~(\cite{iscape}). Data were collected from two Living Lab Stations situated in Guildford, UK, a region characterized by significant vehicle-induced pollution. The monitoring stations, Sutherland Memorial Park (DS\_TS\_099; June--October 2019) and Stoke Park (DS\_TS\_103; February--September 2019), provided hourly open-road measurements.

The primary feature space comprised concentrations of key pollutants (PM$_{2.5}$, CO, NO$_2$, O$_3$) alongside critical atmospheric variables (air temperature, relative humidity, and atmospheric pressure). To enable the neural network architectures to effectively capture the inherent cyclicity of environmental data, temporal features were mathematically transformed. Hour, day, and month values were encoded continuously using sine and cosine functions. Furthermore, to provide the models with spatial awareness and distinguish between the data origins of the two stations, static geographical features, latitude, longitude, and altitude, were appended to the input vectors.

Exploratory data analysis revealed severe transient anomalies in the Stoke Park dataset, primarily manifesting as extreme PM$_{2.5}$ concentration spikes. Readings reached as high as 7,448 $\mu$g/m$^3$, contrasting starkly with the dataset average of approximately 7.957 $\mu$g/m$^3$. During initial experiments, these extreme outliers induced severe numerical instability and exploding gradients across all architectures. Consequently, a threshold clipping strategy was implemented, removing values exceeding 1,000 $\mu$g/m$^3$. This pre-processing step was sufficient to stabilize the loss functions and ensure convergent learning.

To prevent data leakage and simulate a realistic forecasting scenario, the dataset was partitioned chronologically. The hold-out test set was uniformly defined as the month of July for both stations, providing an unseen, common temporal interval for final evaluation. The remaining months from both stations were concatenated to form the training corpus, which was further subdivided into an 80\% training subset and a 20\% validation subset. To standardize the loss landscape, all features across the training, validation, and test sets were normalized to a range of $[0, 1]$ using min-max scaling, with the scaling parameters fitted strictly on the 80\% training subset.

\subsection{Discrete Wavelet Transform (DWT) Pipeline}
To systematically evaluate the impact of multi-scale feature engineering on the quantum and classical architectures, we generated two parallel input streams: a raw temporal feature set and a signal-decomposed feature set. Environmental time series are notoriously non-stationary. While standard Fourier transforms decompose signals into infinite sine and cosine waves, they are constrained by fixed window sizes and fail to capture transient structural changes. To overcome this, we employed a Discrete Wavelet Transform (DWT), which utilizes basis functions localized in both time and frequency. This dual-localization permits the multi-scale assessment of signal discontinuities and long-term trends without losing temporal context.

A five-level DWT was applied independently to each continuous feature in the time series. The algorithm iteratively passes the signal through quadrature mirror filters:
\begin{itemize}
    \item \textbf{High-pass filters} extract the \textit{detail coefficients}, isolating the high-frequency stochastic noise and rapid variance (e.g., sudden pollution spikes).
    \item \textbf{Low-pass filters} extract the \textit{approximation coefficients}, retaining the underlying low-frequency trends (e.g., seasonal or diurnal cycles).
\end{itemize}

To preserve data efficiency, a downsampling operation by a factor of two was applied after each filtering stage. This ensures the total data volume remains comparable to the original signal length. Ultimately, this 5-level decomposition yields five distinct detail signals and one approximation signal per original feature, effectively expanding the dimensional footprint of the input space six-fold.

The performance of the DWT is highly dependent on the choice of the mother wavelet. Rather than selecting a single wavelet arbitrarily, we performed an optimization sweep across established wavelet families. Following the methodology described by \cite{2022Galvao} and \cite{2021Zucatelli}, the optimal wavelet family for each specific feature was dynamically selected by minimizing the Root Mean Square Error (RMSE) between the original raw signal and its reconstructed mathematical approximation. This rigorous selection ensures maximum fidelity in the decomposed latent space utilized by the subsequent neural architectures.

\subsection{Classical Baselines (MLP and LSTM) Optimization}
To establish a rigorous benchmark for the Quantum Neural Network (QNN) architectures, we implemented and optimized two standard classical deep learning models: a Multilayer Perceptron (MLP) and a Long Short-Term Memory (LSTM) network. Both models were implemented using the TensorFlow/Keras framework and were designed to process the exact same input feature space, denoted here by the feature dimension $N_{in}$, which corresponds to the qubit count in the quantum models, and predict the same forecasting horizon ($N_{out}$). The complete architectural specifications, structural variations, and training configurations for both models are summarized in Table \ref{tab:classical_hyperparameters}.

The MLP was constructed as a sequential, fully connected feedforward neural network designed to map the flat input vector of shape $(N_{in},)$ to the forecasting horizon. The architecture comprises two intermediate hidden layers utilizing a sigmoid and a Rectified Linear Unit (ReLU) activation functions, concluding with a Leaky ReLU activation in the output layer. This compact architecture results in a highly constrained hypothesis space, containing a total of 2,025 trainable parameters.

Conversely, to explicitly model the temporal sequence dynamics of the air pollution data, the LSTM network was configured to process temporal windows with an input shape of $(1, N_{in})$. The recurrent layer employs 64 hidden units with Sigmoid activations for both the recurrent state and output gates. Due to its significantly larger parameter space, 61,885 trainable parameters, a Dropout layer with a rate of $0.4$ was integrated immediately following the recurrent operations to provide necessary regularization and prevent overfitting on the training distribution.

To ensure a fair and standardized comparative analysis, both classical models were subjected to identical training regimens. As detailed in Table \ref{tab:classical_hyperparameters}, both networks were trained for 300 epochs using the Adam optimization algorithm with a fixed learning rate of $\eta = 0.001$. The training objective was to minimize the Mean Squared Error (MSE) loss function, with the Mean Absolute Error (MAE) simultaneously monitored to assess absolute predictive deviations.

\begin{table}[h]
    \centering
    \footnotesize
    \begin{tabular}{l l l}
        \toprule
        \textbf{Component/Parameter} & \textbf{Multilayer Perceptron (MLP)} & \textbf{Long Short-Term Memory (LSTM)} \\
        \midrule
        Input Shape & $(N_{in},)$ & $(1, N_{in})$ \\
        Hidden Layer 1 & Dense (10 units, Sigmoid) & LSTM (64 units, Sigmoid) \\
        Hidden Layer 2 & Dense (17 units, ReLU)  & Dropout (Rate = 0.4) \\
        Hidden Layer 3 & -- & Dense (17 units, ReLU) \\
        Output Layer & Dense ($N_{out}$ units, LeakyReLU, $\alpha=0.01$) & Dense ($N_{out}$ units, LeakyReLU, $\alpha=0.01$) \\
        \midrule
        Optimizer & Adam & Adam \\
        Learning Rate ($\eta$) & $0.001$ & $0.001$ \\
        Training Epochs & 300 & 300 \\
        Loss Function & Mean Squared Error (MSE) & Mean Squared Error (MSE) \\
        Tracked Metric & Mean Absolute Error (MAE) & Mean Absolute Error (MAE) \\
        \midrule
        Total Trainable Parameters & 2,025 & 61,885 \\
        \bottomrule
    \end{tabular}
    \caption{Detailed configuration of the Multilayer Perceptron and Long Short-Term Memory models utilized as classical benchmarks against the Quantum Neural Network architectures. $N_{in}$ denotes the number of input features (equivalent to the logical qubit count), and $N_{out}$ denotes the prediction horizon.}
    \label{tab:classical_hyperparameters}
\end{table}

\section{Results}
\label{sec:results}

\subsection{Experimental Performance Analysis}

To resolve the computational bottlenecks inherent to current quantum hardware, we developed a hybrid architecture that couples a gate-based QNN with neuromorphic quantum annealing to forecast PM$_{2.5}$ concentrations up to six hours ahead. A central innovation of this framework is its ability to transcend NISQ-era scalability limits. As illustrated in Figure \ref{fig:qkeraslayer}, by mapping the feature map and ansatz to a Quadratic Unconstrained Binary Optimization (QUBO) problem, our system utilizes the Dynex neuromorphic platform to locate the ground state of the Hamiltonian. This architecture enables the training of large-scale quantum models, up to 172 qubits, leveraging the scalability of neuromorphic annealing for optimization tasks typically intractable on near-term gate-based hardware.

We validated this architecture on high-density urban data from the iSCAPE project (Guildford, UK), enabling the model to learn from real-world open-road conditions characterized by extreme PM$_{2.5}$ concentration spikes. By training parallel models on raw temporal features versus signals decomposed via a five-level Discrete Wavelet Transform (DWT), we investigated a fundamental divergence in how quantum and classical models handle data volatility.

In raw unprocessed inputs, our classical baselines (MLP and LSTM) failed to extract meaningful patterns from the particulate spikes. In stark contrast, the QNN architecture demonstrated an intrinsic resilience to input volatility, successfully capturing the underlying dynamics even without signal decomposition. However, the true advantage of the scalable architecture emerged when integrating DWT. Because wavelet decomposition expands the input feature space six-fold, it necessitates a proportional increase in qubit capacity, which our emulation framework readily accommodates.

Our evaluation reveals two distinct performance regimes governed by data complexity.
In the ``raw data'' regime, Table \ref{tab:base}, the MLP and LSTM classical model baselines failed to converge on stable representations, yielding NMSE values $> 1.2$ and weak correlations ($\rho < 0.35$). In contrast, the QNNs consistently outperformed these baselines across all lookback periods. Notably, the depth-4 QNN achieved a Pearson correlation of $\rho=0.599$ on raw data, effectively doubling the predictive capability of the LSTM ($\rho=0.288$).

The integration of DWT features significantly improved the performance for all model types, narrowing the gap between classical and quantum approaches, Table \ref{tab:wavelets}. However, scaling the quantum architecture yielded substantial dividends in this high-dimensional space. The 172-qubit depth-4 QNN (Wavelet Dataset, Lookback 2) recorded the study's highest average correlation ($\rho=0.729$) and lowest mean error (NMSE = 0.488) across the full 6-hour horizon. Interestingly, the data reveals a non-linear relationship between circuit depth and performance: on Lookback 1 with DWT, Depth 3 achieved the optimal balance ($\rho=0.760$), suggesting that excessive parametrization (Depth 4) without sufficient temporal context (Lookback 2) may lead to diminishing returns or overfitting in the quantum landscape.

\begin{table*}[tp]
    \centering
    \begin{tabular}{cccc|cccc}
        \toprule
        \multicolumn{8}{c}{\textbf{Base Data}} \\
        \toprule
        \multicolumn{4}{c}{Lookback 1 - 16 qubits} & \multicolumn{4}{c}{Lookback 2 - 32 qubits} \\
        \toprule
        Ansatz & NMSE &&  R (Pearson) & Ansatz & NMSE &&  R (Pearson) \\
        \midrule
        QNN depth 1 & 0.784 && 0.551 & QNN depth 1 & 1.005 && 0.485 \\
        QNN depth 2 & \textbf{0.696} && 0.589 & QNN depth 2 & 1.188 && 0.513 \\
        QNN depth 3 & 0.973 && 0.570 & QNN depth 3 & 1.068 && 0.539 \\
        QNN depth 4 & 1.082 && \textbf{0.599} & QNN depth 4 & \textbf{0.907} && \textbf{0.550} \\
        \midrule
        MLP & 1.236 &&  0.303 & MLP & 1.641 &&  0.294 \\
        LSTM & 1.299 &&  0.288 & LSTM & 1.322 &&  0.310 \\
        \bottomrule
    \end{tabular}
    \caption{Comparison of the statistical analyses (NMSE and Pearson's R) for
    the best results of the particulate matter forecasting by the QNN model
    and the Classical Baseline models on the ``raw'' Dataset. Best values for each metric are highlighted in bold.}
    \label{tab:base}
\end{table*}

\begin{table*}[tp]
    \centering
    \begin{tabular}{cccc|cccc}
        \toprule
        \multicolumn{8}{c}{\textbf{Discrete Wavelets Transform}} \\
        \toprule
        \multicolumn{4}{c}{Lookback 1 - 86 qubits} & \multicolumn{4}{c}{Lookback 2 - 172 qubits} \\
        \toprule
        Ansatz & NMSE &&  R (Pearson) & Ansatz & NMSE &&  R (Pearson) \\
        \midrule
        QNN depth 1 & 0.506 && 0.748 & QNN depth 1 & 0.644 && 0.688 \\
        QNN depth 2 & 0.636 && 0.690 & QNN depth 2 & 0.546 && 0.708 \\
        QNN depth 3 & \textbf{0.496} && \textbf{0.760} & QNN depth 3 & 0.754 && 0.581 \\
        QNN depth 4 & 0.545 && 0.708 & QNN depth 4 & \textbf{0.487} && \textbf{0.729} \\
        \midrule
        MLP & \textbf{0.496} &&  0.718 & MLP & 0.589 &&  0.695 \\
        LSTM & 0.597 &&  0.662 & LSTM & 0.563 &&  0.703 \\
        \bottomrule
    \end{tabular}    
    \caption{Comparison of the statistical analyses (NMSE and Pearson's R) for
    the best results of the particulate matter forecasting by the QNN model
    and the Classical Baseline models on the Wavelets Dataset. Best values for each metric are highlighted in bold.}
    \label{tab:wavelets}
\end{table*}

\begin{figure}
    \centering
    \includegraphics[width=\textwidth]{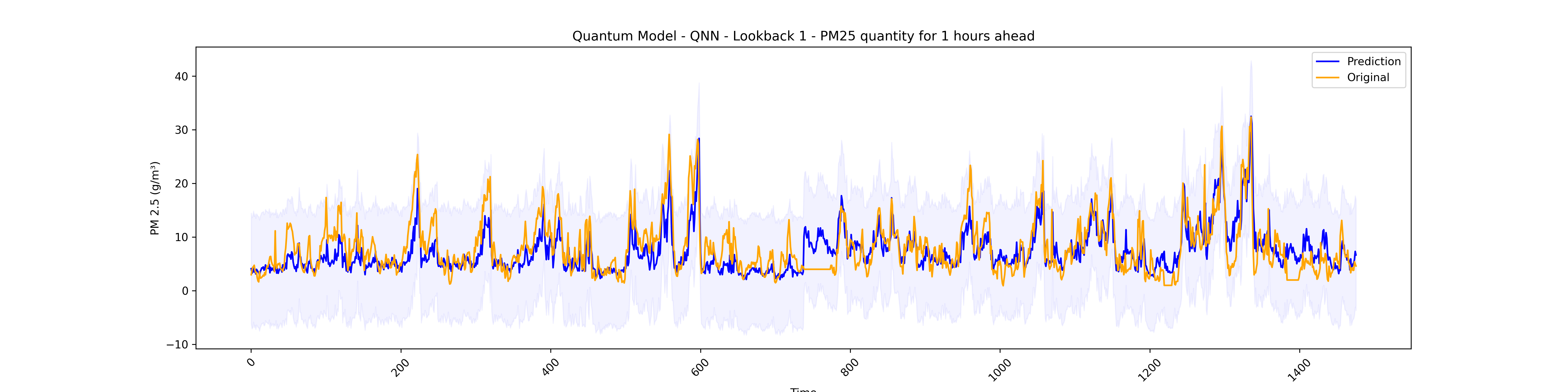}
    \includegraphics[width=\textwidth]{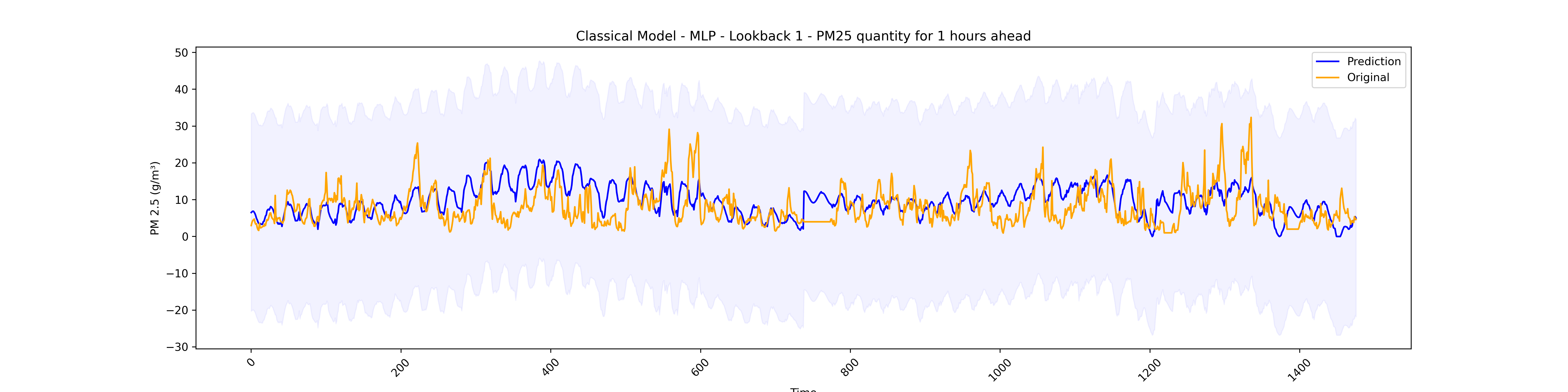}
    \includegraphics[width=\textwidth]{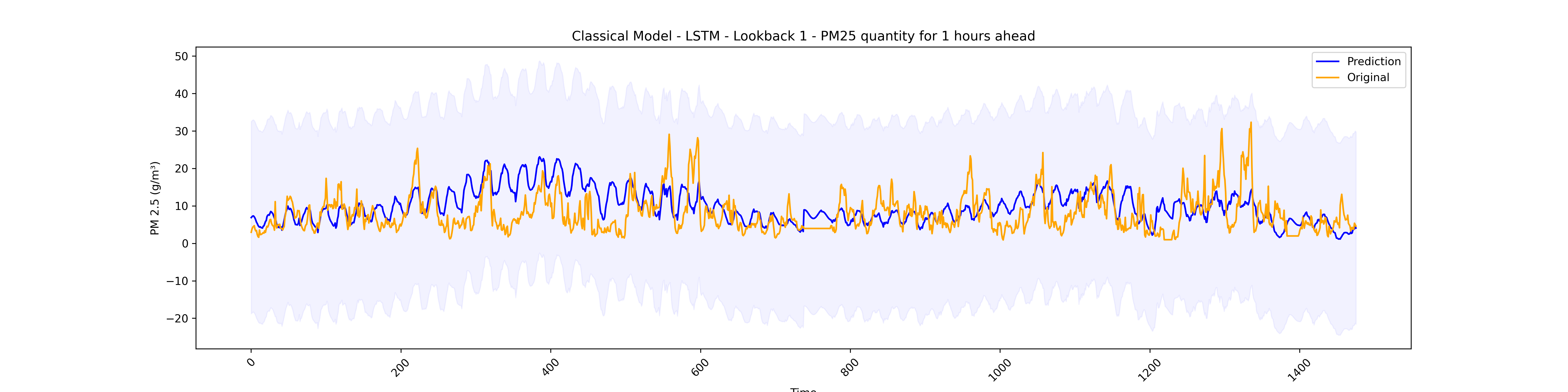}
    \caption{Time series plots illustrating the performance of the proposed QNN model with depth 2 and lookback 1 in forecasting PM$_{2.5}$ concentrations (g/m$^3$) in comparison with the classical baseline models. These results were obtained by training the models with the base dataset, without wavelets. From top to bottom, QNN, MLP, LSTM. In all panels, the blue line represents the model's prediction, and the orange line represents the original (observed) PM$_{2.5}$ data from the test set. The shaded light blue region indicates the 95\% prediction interval, derived using quantile regression, representing the range within which future predicted values are expected to fall with 95\% probability.
    }
    \label{fig:pm25-best}
\end{figure}

Beyond point accuracy, the QNN demonstrated superior uncertainty quantification. Figure \ref{fig:pm25-best} visualizes the model’s 95\% prediction intervals derived from quantile regression. Unlike the classical baselines, which often produced overly confident but erroneous predictions, the QNN’s uncertainty bounds dynamically expanded to encompass extreme spikes.

\subsection{Statistical Significance Analysis}

To rigorously validate the performance variations observed across architectural depths and between quantum and classical paradigms, we conducted a comprehensive non-parametric statistical evaluation of the prediction errors. Initial assessments using the Shapiro-Wilk test verified that the error distributions deviated significantly from normality across all configurations. For instance, testing the 172-qubit depth-4 QNN errors yielded highly significant rejections of the null hypothesis ($W \in [0.533, 0.955]$, $p \ll 0.05$, peaking at $p = 1.20 \times 10^{-121}$), firmly establishing the necessity for non-parametric analysis.

Given the non-normal distributions, a Kruskal-Wallis $H$-test was applied to assess multi-group variance. As summarized in Table \ref{tab:statistical_summary}, the test revealed definitively distinct error populations ($p = 0.000$) across all forecasting horizons when comparing both varying QNN depths and classical baselines. 

To isolate the source of this divergence and confirm quantum advantage, pairwise Wilcoxon Signed-Rank tests were executed. The matrix confirmed that the performance leaps between quantum depths are highly statistically significant. Most crucially, in the most complex feature space (Wavelet, Lookback 2), the depth-4 QNN established absolute statistical superiority ($p = 0.000$) against all other variations and classical baselines across the entirety of the 6-hour horizon. This confirms that the observed state-of-the-art fidelity in the 172-qubit model is robust and fundamentally distinct from the predictive behavior of shallower circuits or classical networks.

\begin{table*}[tp]
    \centering
    \footnotesize
    \begin{tabular}{l c c c}
        \toprule
        \textbf{Dataset Configuration} & \textbf{Shapiro-Wilk ($W$ Range)} & \textbf{Kruskal-Wallis ($H$ Range)} & \textbf{Wilcoxon (Best vs. Others)} \\
        \midrule
        Base (Lookback 1) & $[0.643, 0.671]$, $p \ll 0.05$ & $[123.69, 200.16]$, $p < 10^{-26}$ & Depth 4 vs. D2: $p = 1.90 \times 10^{-27}$ \\
        Base (Lookback 2) & $[0.797, 0.986]$, $p \ll 0.05$ & $[5948.33, 12565.91]$, $p = 0.000$ & Depth 4 vs. D3: $p = 5.29 \times 10^{-84}$ \\
        Wavelet (Lookback 1) & $[0.608, 0.912]$, $p \ll 0.05$ & $[11486.70, 15028.95]$, $p = 0.000$ & Depth 3 vs. All: $p = 0.000$ \\
        Wavelet (Lookback 2) & $[0.533, 0.955]$, $p \ll 0.05$ & $[14539.47, 16259.50]$, $p = 0.000$ & Depth 4 vs. All: $p = 0.000$ \\
        \bottomrule
    \end{tabular}
    \caption{Summary of the statistical metrics evaluating the distributions of prediction errors. The Kruskal-Wallis values represent the multi-group comparison across varying quantum depths. The Wilcoxon column details the pairwise $p$-value confirming the statistical dominance of the best-performing depth for each respective configuration.}
    \label{tab:statistical_summary}
\end{table*}

\section{Conclusion and Discussion}
\label{sec:conclusion}

Our results provide empirical evidence that neuromorphic quantum machine learning can transcend the robustness and scalability barriers that have previously confined QML to theoretical toy-problems. The most significant finding of this work is the QNN's superior performance on a complex, real-world forecasting task, particularly when operating on raw, unprocessed temporal data. This finding is significant, as classical deep learning models, including LSTMs specifically designed for time series, failed to produce meaningful predictions from the same raw data. In contrast, the QNN model effectively captured the underlying dynamics, reducing the NMSE and almost doubling the Pearson correlation. This suggests that the quantum annealing-based computation, which frames the problem as an energy minimization, may possess an intrinsic capability for feature extraction and non-linear dynamics modeling that surpasses these standard classical architectures when data is complex and not explicitly pre-processed.

The application of a DWT reinforces this interpretation. While DWT pre-processing substantially improved the performance of all models, it effectively reduced the performance differential between architectures by classically performing the feature decomposition that the QNN appeared to be handling internally. With DWT-processed features, the QNN was closely matched by the classical MLP. This highlights a key potential advantage of QML: an enhanced ability to learn representations from complex, high-dimensional, or noisy raw signals, potentially reducing the need for intricate, domain-specific feature engineering that is often critical for classical models.

Furthermore, this study provides evidence for the scalability of annealing-based QML algorithms for real-world problems. The successful training and execution of models using DWT-processed features involved up to 172 qubits, a scale that represents a significant challenge for current physical quantum hardware in the NISQ era. Physical NISQ devices are often limited by noise, low qubit connectivity, and decoherence, making benchmarks at this scale difficult. The DynexQML framework, by providing large-scale, hardware-agnostic neuromorphic emulation, serves as a critical tool in this context. It allows for the validation and benchmarking of quantum algorithms at a problem scale relevant for practical applications, bypassing current physical hardware limitations. This ability to test algorithmic scalability today is essential for directing efforts and de-risking the development of future-scale quantum applications.

To date, comparative sustainability analyses have been stalled by a complexity mismatch: one cannot meaningfully compare the energy footprint of a massive classical transformer against a 10-qubit "toy" quantum circuit. For a sustainability audit to be valid, the quantum model must first demonstrate the capacity to solve a significant, industrially relevant problem. We argue that the era of speculative "quantum advantage" is ending; with the Dynex-enabled scaling demonstrated here, the field can now transition to measuring the concrete "energy-to-solution" ratio of high-dimensional quantum solvers. Consequently, this framework not only advances forecasting accuracy but also defines the experimental boundary where the sustainability of quantum algorithms can finally be empirically tested against classical deep learning.

While our findings are promising, it is important to acknowledge the limitations of this work. The QNN performance was demonstrated on an emulated platform, and these results do not account for the noise models, limited connectivity, or decoherence inherent to physical quantum devices. Benchmarking this framework on physical hardware is a critical next step to validate the translation of this emulated performance to a true physical quantum advantage. Moreover, the performance and generalizability of the DynexQML framework should be compared against a wider array of state-of-the-art classical models, such as temporal convolutional networks (TCNs) or transformers.

In conclusion, this work demonstrates the practical application and potential advantages of a scalable quantum annealing-based QML framework. The QNN's superior ability to model raw, complex time series data suggests a promising pathway for QML to tackle real-world problems where classical feature extraction is a significant bottleneck. The DynexQML platform's ability to operate at a scale of hundreds of qubits provides an essential testbed for prototyping and validating quantum algorithms, bridging the gap between near-term hardware and the demands of industrially-relevant applications. Future work will focus on translating these findings to physical hardware and exploring the boundaries of QML performance in other complex domains, keeping sustainability at the forefront of algorithm design.

\section*{Declarations}

\textbf{Funding:} This work has been partially supported by QuIIN - EMBRAPII CIMATEC Competence Center in Quantum Technologies, with financial resources from the PPI IoT/Manufatura 4.0 / PPI HardwareBR of the MCTI grant number 053/2023, signed with EMBRAPII. We also thank the National Council for Scientific and Technological Development (CNPq, Brazil), for partially funding this work. Erick G. Sperandio Nascimento and Marcelo A. Moret are CNPq technological development fellows, Grant Numbers Proc. 313527/2026-1 and 305096/2022-2, respectively.

\textbf{Acknowledgements:} The authors would like to thank the Supercomputing Center for Industrial Innovation (CS2I), the Reference Center for Artificial Intelligence (CRIA), and the Latin American Quantum Computing Center (LAQCC), all from SENAI CIMATEC University, for providing technical and infrastructure support. The authors thank the Surrey Institute for People-Centered Artificial Intelligence for its support. The authors also acknowledge Dynex Holding Establishment for providing access to their neuromorphic quantum computing platform and technical expertise in quantum machine learning methodologies. The platform's neuromorphic architecture, protected under international patent WO2024231907, transforms conventional GPUs into quantum processors through memristive emulation, achieving quantum behavior at scale with support for over one million algorithmic qubits. This research utilized Dynex's Quantum Bridge technology, which systematically translates quantum gate-based circuits into quantum annealing formulations, and the DynexQML framework featuring the QKerasLayer for quantum neural networks. Dynex's Quantum-as-a-Service model democratizes access to quantum computing resources as quantum-powered solutions while maintaining cost advantages of up to 90\% compared to traditional quantum systems. This collaboration demonstrates the practical viability of neuromorphic quantum computing for real-world environmental monitoring and prediction tasks.

\textbf{Data Availability:} The data is available through the iSCAPE website. The datasets used were Living Lab Station data Feb-Sept 2019, Stoke Park, Guildford (https://zenodo.org/records/3609183) and Living Lab Station data Jun-Oct 2019, Sutherland Memorial Park, Guildford (https://zenodo.org/records/3609187).

\textbf{Code Availability:} All code used in the experiments is available through the following link: https://github.com/OttoMP/quantum-air-pollution-nowcasting.

\textbf{Authors' Contributions:} Erick Sperandio and Otto Pires conceived the project. Erick Sperandio contributed with the methodology by providing expertise in AI, quantum computing and statistical analysis for air pollution forecasting, along side Otto Pires who provided expertise in quantum computing and AI. Otto Pires and André Saimon coded the quantum circuits and AI models. André Saimon performed the statistical analyses. Otto Pires prepared the final version of the released code and datasets. Daniela Herrmann offered the DynexQML framework and the Quantum Bridge infrastructure for experiments. Otto Pires wrote the manuscript alongside Erick Sperandio, with input from all co-authors. Prashant Kumar provided guidance on the air pollution aspects, as well as feedback on the final version of the manuscript. Erick Sperandio led the research. Marcelo Moret and Erick Sperandio supervised the project.

\textbf{Competing Interests:} Daniela Herrmann is an employee of Dynex. All other authors declare no competing interests.

%Bibliography
\bibliographystyle{unsrt}  
\bibliography{references}

\end{document}